\pdfoutput=1
\documentclass[12pt,a4paper]{article}

\usepackage{ifthen} 
\newboolean{pdflatex}
\setboolean{pdflatex}{true} 

\newboolean{articletitles}
\setboolean{articletitles}{true} 

\newboolean{uprightparticles}
\setboolean{uprightparticles}{false} 

\def\paperauthors{LHCb collaboration} 
\def\paperasciititle{Energy-energy correlators within Z-tagged jets in proton-proton collisions at \sqrt{s} = 13 TeV} 
\def\papertitle{Energy-energy correlators within $Z$-tagged jets in proton-proton collisions at $\sqrt{s}$ = 13 Te\hspace{-2pt}V} 
\def\paperkeywords{{High Energy Physics}, {LHCb}} 
\def\papercopyright{\the\year\ CERN for the benefit of the LHCb collaboration} 
\def\paperlicence{CC BY 4.0 licence}
\def\paperlicenceurl{https://creativecommons.org/licenses/by/4.0/}

\newif\ifEnableSectionTOCLinks
\EnableSectionTOCLinksfalse 

\usepackage[top=1in, bottom=1.25in, left=1in, right=1in]{geometry}

\usepackage{microtype}
\usepackage{lineno}  
\usepackage{xspace} 
\usepackage{caption} 

\usepackage{graphicx}  
\usepackage{color}
\usepackage{colortbl}
\graphicspath{{./figs/}} 

\usepackage{amsmath} 
\usepackage{amssymb}
\usepackage{amsfonts}
\usepackage{upgreek} 

\usepackage[normalem]{ulem} 

\newcommand*\patchAmsMathEnvironmentForLineno[1]{%
\expandafter\let\csname old#1\expandafter\endcsname\csname #1\endcsname
\expandafter\let\csname oldend#1\expandafter\endcsname\csname
end#1\endcsname
 \renewenvironment{#1}%
   {\linenomath\csname old#1\endcsname}%
   {\csname oldend#1\endcsname\endlinenomath}%
}
\newcommand*\patchBothAmsMathEnvironmentsForLineno[1]{%
  \patchAmsMathEnvironmentForLineno{#1}%
  \patchAmsMathEnvironmentForLineno{#1*}%
}
\AtBeginDocument{%
\patchBothAmsMathEnvironmentsForLineno{equation}%
\patchBothAmsMathEnvironmentsForLineno{align}%
\patchBothAmsMathEnvironmentsForLineno{flalign}%
\patchBothAmsMathEnvironmentsForLineno{alignat}%
\patchBothAmsMathEnvironmentsForLineno{gather}%
\patchBothAmsMathEnvironmentsForLineno{multline}%
\patchBothAmsMathEnvironmentsForLineno{eqnarray}%
}

\usepackage[pdftex,
            pdfauthor={\paperauthors},
            pdftitle={\paperasciititle},
            pdfkeywords={\paperkeywords}]{hyperref}
\usepackage{hyperxmp}
\hypersetup{
    pdfcopyright={Copyright (C) \papercopyright},
    pdflicenseurl={\paperlicenceurl}
}

\usepackage[colorinlistoftodos,textsize=scriptsize]{todonotes}

\usepackage[bottom,flushmargin,hang,multiple]{footmisc}

\usepackage[all]{hypcap} 

\usepackage{xspace} 
\usepackage{upgreek}

\def\lhcb   {\mbox{LHCb}\xspace}

\def\lhc    {\mbox{LHC}\xspace}

\def\MagUp {\mbox{\em Mag\kern -0.05em Up}\xspace}

\ifdefined\ifuprightparticles
\else
\newboolean{uprightparticles}
\setboolean{uprightparticles}{false} 
\fi

\ifthenelse{\boolean{uprightparticles}}%
{

 \def\PDelta      {\ensuremath{\Delta}\xspace}                 
 \def\PXi         {\ensuremath{\Xi}\xspace}                 
 \def\PLambda     {\ensuremath{\Lambda}\xspace}                 
 \def\PSigma      {\ensuremath{\Sigma}\xspace}                 
 \def\POmega      {\ensuremath{\Omega}\xspace}                 
 \def\PUpsilon    {\ensuremath{\Upsilon}\xspace}
 \let\oldPi\Pi
 \def\PPi         {\ensuremath{\oldPi}\xspace}

 \def\PB      {\ensuremath{\mathrm{B}}\xspace}                 
 \def\PD      {\ensuremath{\mathrm{D}}\xspace}                 

 \def\PK      {\ensuremath{\mathrm{K}}\xspace}                 
 \def\PW      {\ensuremath{\mathrm{W}}\xspace}                 
 \def\PZ      {\ensuremath{\mathrm{Z}}\xspace}                 

 \def\Pe      {\ensuremath{\mathrm{e}}\xspace}                 
 \def\Pp      {\ensuremath{\mathrm{p}}\xspace}                 

 \def\Ps      {\ensuremath{\mathrm{s}}\xspace}

 \def\thebaroffset{0.0em}
}
{

 \mathchardef\PDelta="7101
 \mathchardef\PXi="7104
 \mathchardef\PLambda="7103
 \mathchardef\PSigma="7106
 \mathchardef\POmega="710A
 \mathchardef\PUpsilon="7107
 \mathchardef\PPi="7105
 \def\PB      {\ensuremath{B}\xspace}                 
 \def\PD      {\ensuremath{D}\xspace}                 

 \def\PK      {\ensuremath{K}\xspace}                 
 \def\PW      {\ensuremath{W}\xspace}                 
 \def\PZ      {\ensuremath{Z}\xspace}                 

 \def\Pe      {\ensuremath{e}\xspace}                 
 \def\Pp      {\ensuremath{p}\xspace}                 

 \def\Ps      {\ensuremath{s}\xspace}

 \def\thebaroffset{0.18em}
}
\newcommand{\offsetoverline}[2][\thebaroffset]{\kern #1\overline{\kern -#1 #2}}%

\makeatletter
\ifcase \@ptsize \relax
  \newcommand{\miniscule}{\@setfontsize\miniscule{4}{5}}
\or
  \newcommand{\miniscule}{\@setfontsize\miniscule{5}{6}}
\or
  \newcommand{\miniscule}{\@setfontsize\miniscule{5}{6}}
\fi
\makeatother

\DeclareRobustCommand{\optbar}[1]{\shortstack{{\miniscule (\rule[.5ex]{1.25em}{.18mm})}
  \\ [-.7ex] $#1$}}

\def\epem       {{\ensuremath{\Pe^+\Pe^-}}\xspace}

\def\W      {{\ensuremath{\PW}}\xspace}

\def\Z      {{\ensuremath{\PZ}}\xspace}

\def\squark    {{\ensuremath{\Ps}}\xspace}

\def\KorKbar {\kern \thebaroffset\optbar{\kern -\thebaroffset \PK}{}\xspace}

\def\D       {{\ensuremath{\PD}}\xspace}

\def\DorDbar {\kern \thebaroffset\optbar{\kern -\thebaroffset \PD}\xspace}

\def\Dp      {{\ensuremath{\D^+}}\xspace}
\def\Dm      {{\ensuremath{\D^-}}\xspace}

\def\DpDm    {\ensuremath{\Dp {\kern -0.16em \Dm}}\xspace}

\def\B       {{\ensuremath{\PB}}\xspace}

\def\BorBbar {\kern \thebaroffset\optbar{\kern -\thebaroffset \PB}\xspace}

\def\Bd      {{\ensuremath{\B^0}}\xspace}

\def\BdorBdbar {\kern \thebaroffset\optbar{\kern -\thebaroffset \Bd}\xspace}

\def\Bs      {{\ensuremath{\B^0_\squark}}\xspace}

\def\BsorBsbar {\kern \thebaroffset\optbar{\kern -\thebaroffset \Bs}\xspace}

\def\Y#1S{\ensuremath{\PUpsilon{(#1S)}}\xspace}

\def\proton      {{\ensuremath{\Pp}}\xspace}

\def\LorLbar     {\kern \thebaroffset\optbar{\kern -\thebaroffset \PLambda}\xspace}

\def\to                 {\ensuremath{\rightarrow}\xspace}

\newcommand{\lqcd}{{\ensuremath{\Lambda_{\mathrm{QCD}}}}\xspace}

\def\AT#1     {\ensuremath{A_{\mathrm{T}}^{#1}}\xspace}           

\def\C#1      {\ensuremath{\mathcal{C}_{#1}}\xspace}                       
\def\Cp#1     {\ensuremath{\mathcal{C}_{#1}^{'}}\xspace}                    
\def\Ceff#1   {\ensuremath{\mathcal{C}_{#1}^{\mathrm{(eff)}}}\xspace}        
\def\Cpeff#1  {\ensuremath{\mathcal{C}_{#1}^{'\mathrm{(eff)}}}\xspace}       
\def\Ope#1    {\ensuremath{\mathcal{O}_{#1}}\xspace}                       
\def\Opep#1   {\ensuremath{\mathcal{O}_{#1}^{'}}\xspace}                    

\newcommand{\aunit}[1]{\ensuremath{\text{\,#1}}}       

\newcommand{\tev}{\aunit{Te\kern -0.1em V}\xspace}
\newcommand{\gev}{\aunit{Ge\kern -0.1em V}\xspace}
\newcommand{\mev}{\aunit{Me\kern -0.1em V}\xspace}
\newcommand{\kev}{\aunit{ke\kern -0.1em V}\xspace}
\newcommand{\ev}{\aunit{e\kern -0.1em V}\xspace}
 
\newcommand{\mevc}{\ensuremath{\aunit{Me\kern -0.1em V\!/}c}\xspace}
\newcommand{\gevc}{\ensuremath{\aunit{Ge\kern -0.1em V\!/}c}\xspace}
\newcommand{\mevcc}{\ensuremath{\aunit{Me\kern -0.1em V\!/}c^2}\xspace}
\newcommand{\gevcc}{\ensuremath{\aunit{Ge\kern -0.1em V\!/}c^2}\xspace}

\def\fb   {\ensuremath{\aunit{fb}}\xspace}
\def\invfb   {\ensuremath{\fb^{-1}}\xspace}

\def\deriv {\ensuremath{\mathrm{d}}}

\def\gsim{{~\raise.15em\hbox{$>$}\kern-.85em
          \lower.35em\hbox{$\sim$}~}\xspace}
\def\lsim{{~\raise.15em\hbox{$<$}\kern-.85em
          \lower.35em\hbox{$\sim$}~}\xspace}

\def\sqs   {\ensuremath{\protect\sqrt{s}}\xspace}

\def\pt         {\ensuremath{p_{\mathrm{T}}}\xspace}

\def\evtgen     {\mbox{\textsc{EvtGen}}\xspace}

\def\geant      {\mbox{\textsc{Geant4}}\xspace}

\def\photos     {\mbox{\textsc{Photos}}\xspace}

\def\pythia     {\mbox{\textsc{Pythia}}\xspace}

\def\tell1  {TELL1\xspace}
\def\ukl1   {UKL1\xspace}

\newcommand{\eg}{\mbox{\itshape e.g.}\xspace}

\newcommand{\lhcborcid}[1]{\href{https://orcid.org/#1}{\hspace*{0.1em}\raisebox{-0.45ex}{\includegraphics[width=1em]{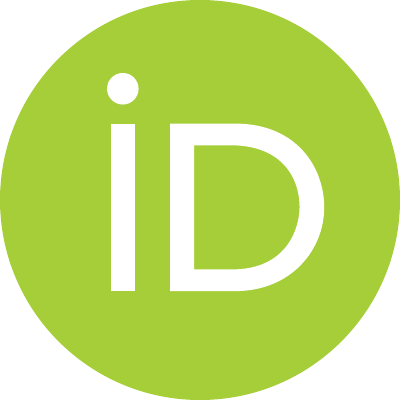}}}}

\hypersetup{
  colorlinks   = true, 
  urlcolor     = blue, 
  linkcolor    = blue, 
  citecolor    = red   
}

\ifEnableSectionTOCLinks
    \usepackage[explicit]{titlesec} 
    
    \let\oldcontentsline\contentsline
    \renewcommand

    \titleformat{\section}{\normalfont\Large\bf}{\hyperlink{tocsection.\thesection}{{\thesection} \parbox[t]{\dimexpr\textwidth-1pc}{#1}}}{1pc}{}

    \titleformat{\subsection}{\normalfont\bf}{\hyperlink{tocsubsection.\thesubsection}{{\thesubsection} \parbox[t]{\dimexpr\textwidth-1pc}{#1}}}{1pc}{}

    \titleformat{name=\section,numberless}[display]{}{}{0pt}{\normalfont\Huge\bfseries #1}
\fi

\usepackage{cite} 
\usepackage{LHCb/mciteplus}
\makeatletter
\g@addto@macro\bfseries{\boldmath}
\makeatother

\usepackage{longtable} 

\begin{document}

\renewcommand{\thefootnote}{\fnsymbol{footnote}}
\setcounter{footnote}{1}


\begin{titlepage}
\pagenumbering{roman}

\vspace*{-1.5cm}
\centerline{\large EUROPEAN ORGANIZATION FOR NUCLEAR RESEARCH (CERN)}
\vspace*{1.5cm}
\noindent
\begin{tabular*}{\linewidth}{lc@{\extracolsep{\fill}}r@{\extracolsep{0pt}}}
\ifthenelse{\boolean{pdflatex}}
{\vspace*{-1.5cm}\mbox{\!\!\!\includegraphics[width=.14\textwidth]{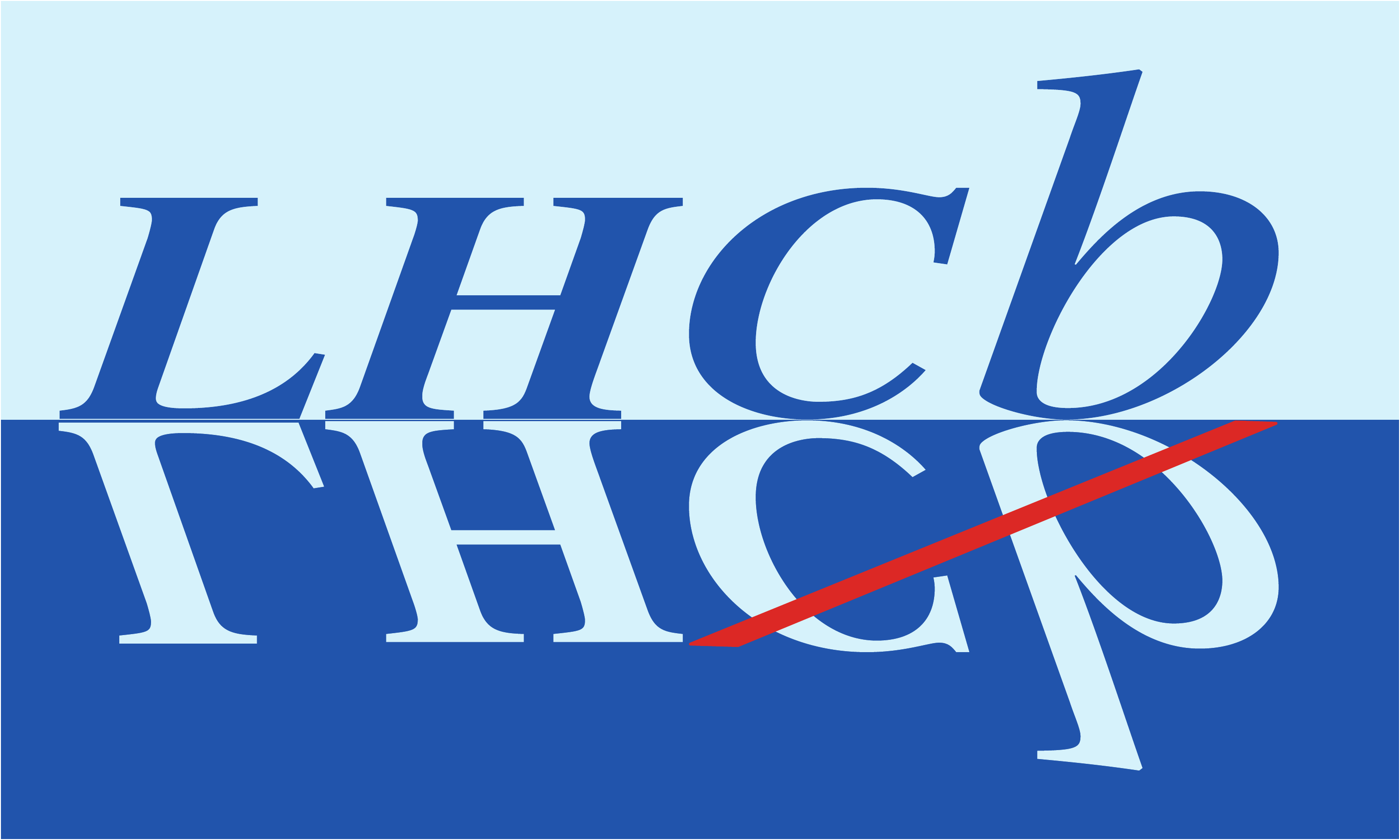}} & &}%
{\vspace*{-1.2cm}\mbox{\!\!\!\includegraphics[width=.12\textwidth]{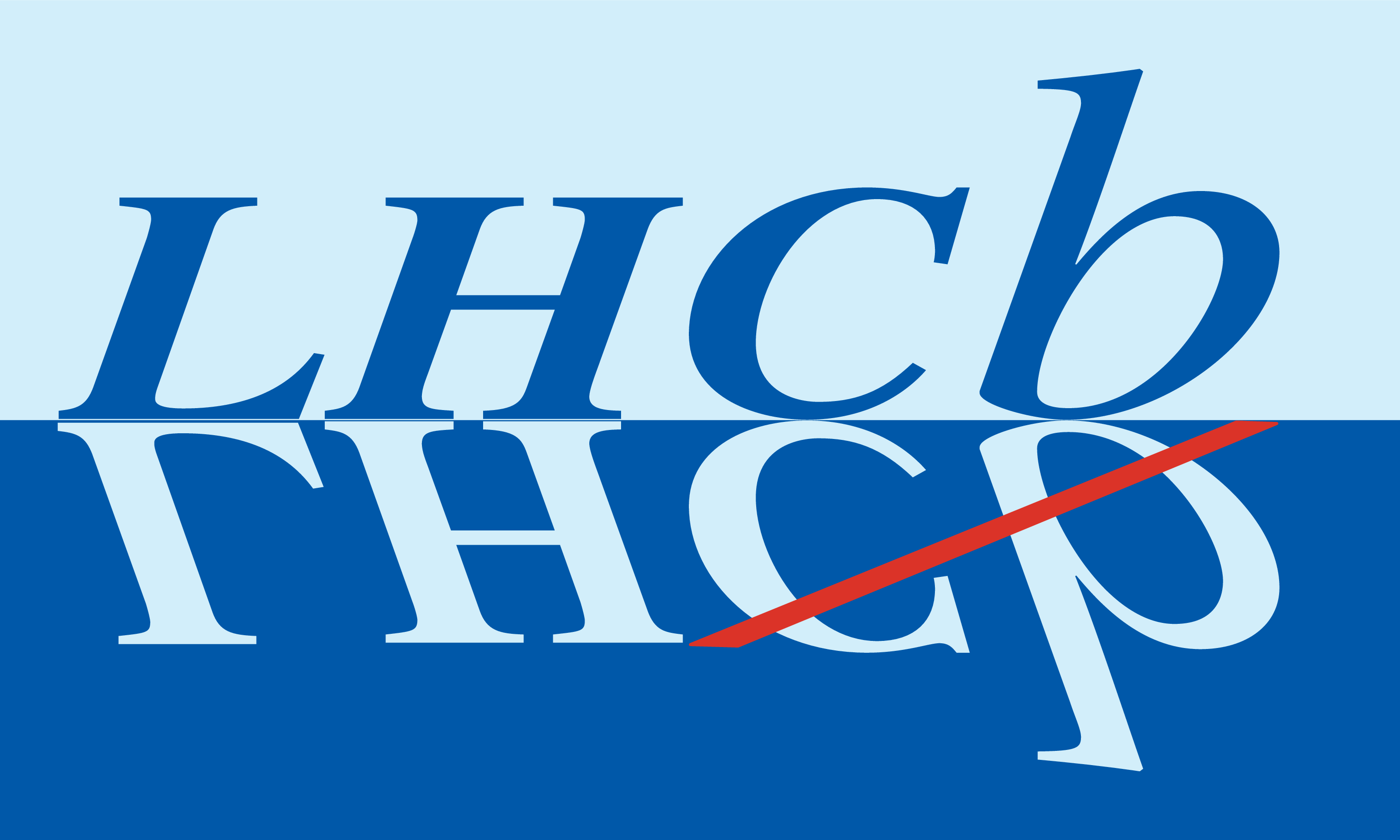}} & &}%
\\
 & & CERN-EP-2026-257 \\  
 & & LHCb-PAPER-2026-032 \\  
 & & October 9, 2026 \\ 
 & & \\
\end{tabular*}

\vspace*{4.0cm}

{\normalfont\bfseries\boldmath\huge
\begin{center}
  \papertitle 
\end{center}
}

\vspace*{2.0cm}

\begin{center}
\paperauthors\footnote{Authors are listed at the end of this Letter.}
\end{center}

\vspace{\fill}

\begin{abstract}
  \noindent
  Energy-energy correlators probe perturbative and nonperturbative QCD regimes of jet formation through momentum-weighted angular correlations between hadrons. First measurements in light-quark-enriched jets are presented for inclusive, like-sign, and opposite-sign charged-hadron pairs, using proton-proton collisions at LHCb. The results support universality of the confinement scale for light-quark-initiated jets across collision systems and the reported transverse momentum scale of the jet. In the charge-dependent measurements, opposite-sign correlations are enhanced as angular separation decreases down to the confinement-scale region, a behavior that is well described by simulations.
\end{abstract}

\vspace*{2.0cm}

\begin{center}
  Submitted to Phys.~Rev.~Lett.
\end{center}

\vspace{\fill}

{\footnotesize 
\centerline{\copyright~\papercopyright. \href{\paperlicenceurl}{\paperlicence}.}}
\vspace*{2mm}

\end{titlepage}


\newpage
\setcounter{page}{2}
\mbox{~}
%
%
%
%


\renewcommand{\thefootnote}{\arabic{footnote}}
\setcounter{footnote}{0}


\cleardoublepage


\pagestyle{plain} 
\setcounter{page}{1}
\pagenumbering{arabic}


Energy correlators in high-energy particle physics were introduced more than 40 years ago to describe energy flow patterns in the products of high-energy electron-positron annihilation~\cite{Basham:1977iq, Basham:1978bw, Basham:1979gh}.  With the advent of the Large Hadron Collider (\lhc), there has been a resurgence of interest in energy correlators as versatile observables to predict and study a wide range of multiparticle correlations in hadron-hadron collisions~\cite{Hofman:2008ar, Chen:2020vvp, Lee:2022uwt,Moult:2025nhu}, as energy correlators encode information about the stages of jet formation in the energy-weighted angular distributions of final-state particles. This resurgence has prompted a diverse program of measurements that have been performed for two-point and three-point energy correlators by the ALICE and CMS experiments at the LHC~\cite{CMS:2024mlf, ALICE:2024dfl, CMS:2025ydi, ALICE:2025igw, ALICE:2026zyx}, and by the STAR experiment at the Relativistic Heavy Ion Collider (RHIC)~\cite{STAR:2025jut}. Energy-correlator measurements have so far focused on particle correlations within jets, collimated sprays of particles produced by an outgoing high-energy quark or gluon~\cite{Hanson:1975fe}. Jets provide an excellent environment to study the strong interaction, described by quantum chromodynamics~(QCD), since their internal composition can shed light on the dynamics of the jet formation process related to different regimes of QCD~\cite{Kogler:2018hem}.

In this Letter, two-point energy correlators are measured for the first time in jets produced back-to-back with a \Z boson for inclusive, like-sign, and opposite-sign pairs of charged hadrons.  As forward-rapidity \Z+ jet production predominantly consists of light-quark-initiated jets, the reported measurements offer a timely first look into the two-point energy correlators of such jets in \proton\proton collisions, considering that recent studies propose two-point energy correlators to discriminate the initiating parton of jets~\cite{Apolinario:2025vtx, Chen:2024quk, Herrmann:2025fqy} and to test universality of nonperturbative physics~\cite{Liu:2024lxy,Barata:2026uby}.

The emergence of regions associated with specific stages of the jet formation process~\cite{Komiske:2022enw} is a fundamental feature of two-point energy correlators in jets. The \textit{perturbative region} at large angular separations is associated with earlier times in the formation of the jet, when parton showering occurs, and is suitable for perturbative QCD (pQCD) predictions. The \textit{free-hadron region} at small values of angular separation is associated with later times in the formation of the jet. At these small angular separations, hadrons have already fully formed and predominantly propagate without interacting with each other. This region is suitable to study nonperturbative aspects of QCD, such as hadronization models, given that the kinematic distributions of produced hadrons can be sensitive to these models (see \eg~Ref.~\cite{Webber:1999ui}). The \textit{transition region} is defined by the change of scaling behavior between the perturbative and free-hadron regions and indicates the onset of hadronization. In the case of light-quark- or gluon-initiated jets, the transition region corresponds to angular separations on the order of $\lqcd$ divided by the transverse momentum of the jet with respect to the proton-proton (\proton\proton) collision axis, $p_{\mathrm{T,\,jet}}$~\cite{Komiske:2022enw, Craft:2022kdo}, with light-quark-initiated jets predicted to have a transition region at slightly smaller angular separation than gluon-initiated jets~\cite{Apolinario:2025vtx}.

The two-point energy correlators can also be used to study jet evolution dynamics through two-particle configurations with specific quantum numbers, such as the electric charge or spin~\cite{Lee:2023npz}. Given that different angular-separation regions are correlated with different stages of the jet formation process, studying charge-dependent energy correlators in jets can reveal charge-dependent dynamics in hadron-pair production sensitive to different hadronization models~\cite{Andersson:1983ia,Webber:1983if}. Therefore, measuring the inclusive and charge-dependent energy correlators in the transition and free-hadron regions provides a means to probe the hadronization process, which cannot be described using perturbative calculation techniques. 
The two-point energy correlator measured in this Letter, also called the energy-energy correlator~(EEC), is defined as a yield of momentum-weighted charged-hadron pairs in jets produced back-to-back with a \Z boson. The EECs are presented as a function of $p_{\mathrm{T,\,jet}}$ and the angular distance $R_{\textrm{L}}$ in rapidity $y$ and azimuthal angle $\phi$,
\begin{equation}
    R_{\textrm{L}}=\sqrt{(\Delta y)^2 + (\Delta \phi)^2},
\end{equation}
where the difference in azimuthal angle $\Delta\phi$ is measured in the plane perpendicular to the beam line. Thus, the inclusive EEC is defined as
\begin{equation}\label{eq:eec-def}
\begin{split}
    \Sigma_{\text{EEC}}(R_{\textrm{L}},p_{\mathrm{T,\,jet}}) &= \frac{1}{N_{Z+\textrm{jets}}\cdot\Delta_{R_{\textrm{L}}}}\times \\ &\:\sum_{\{Z+\textrm{jets}\}}\sum_{i<j}\int_{\Delta_{R_{\textrm{L}}}} w^{ij}_{\pt} \delta(R'_\textrm{L}-R^{ij}_{\textrm{L}})\deriv R'_\textrm{L} ,
\end{split}
\end{equation}
where $N_{Z+\textrm{jets}}$ is the number of \Z-tagged jets, and $\Delta_{R_{\textrm{L}}}$ refers to the width of the corresponding $R_{\textrm{L}}$ interval (bin). The sums run over all the pairs inside the selected \Z-tagged jets, where every pair formed by charged hadrons $i$ and $j$ has an angular distance $R^{ij}_L$, and is weighted by their respective transverse momenta, $p^{i}_{\mathrm{T}}$ and $p^{j}_{\mathrm{T}}$ by
\begin{equation}
    w^{ij}_{\pt} = \frac{p^i_{\mathrm{T}} p^j_{\mathrm{T}}}{(p_{\mathrm{T,\,jet}})^2},
\end{equation}
where $w^{ij}_{\pt}$ is the momentum weight. 

The charge-dependent EECs are defined as the difference between like-sign (LS) EECs, $\Sigma^{\text{LS}}_{\text{EEC}}$, and opposite-sign (OS) EECs, $\Sigma^{\text{OS}}_{\text{EEC}}$. The LS EECs are built with like-sign charged hadrons, and the OS EECs are built with opposite-sign charged hadrons, where both quantities are normalized by the inclusive EECs.

In this Letter, the inclusive and charge-dependent EECs as a function of $(R_{\textrm{L}}, p_{\mathrm{T,\,jet}})$, and the inclusive EECs as a function of $(R_{\textrm{L}} \langle p_{\mathrm{T,\,jet}}\rangle, p_{\mathrm{T,\,jet}})$, where $\langle p_{\mathrm{T,\,jet}}\rangle$ is the average of the jet transverse momentum, are reported. The \proton\proton collision data collected between 2016 and 2018 by the LHCb detector and corresponding to a total integrated luminosity of $5.4\invfb$ are used. The \lhcb detector is a single-arm forward spectrometer covering the pseudorapidity range $2 < \eta < 5$, described in detail in Refs.~\cite{LHCb-DP-2008-001, LHCb-DP-2014-002}. The detector elements that are particularly relevant to this measurement are the high-precision tracking system~\cite{LHCb-DP-2014-001,LHCb-DP-2017-001} and the muon detection system\cite{LHCb-DP-2012-002}.

Charged and neutral particles are clustered into jets with the anti-$k_T$ algorithm~\cite{Cacciari:2008gp} using a jet resolution parameter (radius) of $R=0.5$ and $E$-scheme recombination, as described in Ref.~\cite{LHCb-PAPER-2022-013}. Jet identification requirements are applied to reduce the rate of misreconstructed jets and to minimize jet energy corrections. Reconstructed jets are required to be the highest-$p_{\textrm{T}}$ jet in their respective event and to be produced in the opposite direction of the \Z boson ($\Delta\phi_{\Z\text{-jet}} > 7\pi/8$), reconstructed via its dimuon decay channel, which provides a clean measurement environment given its high signal-to-noise ratio~\cite{LHCb-PAPER-2013-058, LHCb-PAPER-2015-049}. The \mbox{\Z + jet} events are predominantly produced by QCD Compton scattering ($g+q\to\Z+q$) due to the asymmetry between the gluon and light-quark parton distribution functions in the \lhcb acceptance, resulting in a jet sample enriched by light-quark-initiated jets~\cite{LHCb-PAPER-2022-013}. The muons produced by the \Z boson decay are required to be outside the selected jet cone, and satisfy $2<\eta_{\textrm{muon}}<4.5$ and $p_{\textrm{T,muon}}>20\gev$ (natural units with $\hbar = c = 1$ are used throughout). The jets have to be in the $20 < p_{\mathrm{T,\,jet}} < 100\gev$ range and their axes must satisfy a rapidity requirement $2.5<y_{\textrm{jet}}<4$ to ensure all constituents lie within the nominal \lhcb acceptance. After the jet selection, the charged hadrons inside jets must be within $2 < y_{\textrm{hadron}} < 4.5$, where the energy used in the rapidity estimation is calculated with the particle momentum and a mass assumption based on the particle identification system. Finally, the charged hadrons must be within a kinematic range given by $\pt > 0.25\gev$ and $4 < p < 1000\gev$, and must meet minimal track quality requirements. 

The corrections applied to the EEC distributions for detector effects are obtained using simulation. The particle-level events are generated using \mbox{\pythia8.186}~\cite{Sjostrand:2007gs} with a specific \lhcb configuration~\cite{LHCb-PROC-2010-056}. The decay of unstable particles is handled by \evtgen~\cite{Lange:2001uf}, and the final-state radiation by \photos~\cite{davidson2015photos}. The interaction of particles with matter is simulated using \geant~\cite{Agostinelli:2002hh,Allison:2006ve} as specified in Ref.~\cite{LHCb-PROC-2011-006}. The output of \geant is referred to as the detector-level simulation sample. A geometrical jet-matching procedure is performed between detector- and particle-level jets to estimate jet reconstruction purities and efficiencies. The angular distance between the jet axes is required to be smaller than the jet resolution parameter $0.5$. In the case of multiple candidates, that which minimizes the angular distance between axes is chosen.

The number of jets is corrected as a function of $p_{\mathrm{T,\,jet}}$. First, efficiency corrections for the muons from the \Z boson decay are implemented as an event weight. Then, the jet reconstruction purity accounts for misreconstructed jets, and it is estimated by the fraction of detector-level jets that have a matching particle-level jet, which ranges between $89\%$ and $98\%$. Afterwards, an unfolding process is performed using the Iterative Bayesian Unfolding (IBU) algorithm~\cite{DAGOSTINI1995487} as implemented in the \textsc{RooUnfold} package~\cite{RooUnfold}; further details on the unfolding process are provided in Appendix~\ref{appendix:unfolding}. Finally, a jet reconstruction efficiency is defined to account for jets that are not reconstructed, estimated by the fraction of particle-level jets with a matching detector-level jet, which ranges between $79\%$ and $97\%$.

The momentum weights play a fundamental role in the measurement of the two-point energy correlators, and their dependence on a significantly smeared variable, such as $p_{\mathrm{T,\,jet}}$, makes their unfolding imperative. Therefore, before unfolding the momentum weights, the number of charged-hadron pairs is corrected and unfolded in a three-dimensional binning of $(R_{\textrm{L}}, p_{\mathrm{T,\,jet}},w_{\pt})$.
In a similar fashion to the case of jets, a pair-matching procedure between detector-level pairs and particle-level pairs is necessary to estimate the reconstruction purities and efficiencies. A pair is successfully matched, or correctly reconstructed, when both constituents in a pair are matched between particle- and detector-level; additionally, it is required that these matched pairs must be in matched jets. 

The number of charged-hadron pairs is corrected as a function of $(R_{\textrm{L}}, p_{\mathrm{T,\,jet}},w_{\pt})$ in a similar fashion to jets. First, pair reconstruction purities account for misreconstructed pairs. Afterwards, an unfolding process is performed using the IBU algorithm simultaneously in three dimensions $(R_{\textrm{L}}, p_{\mathrm{T,\,jet}},w_{\pt})$. Finally, pair reconstruction efficiencies account for pairs that were not reconstructed. The reconstruction purities range between $66\%$ and $74\%$, while the reconstruction efficiencies between $41\%$ and $66\%$, where lower $p_{\mathrm{T,\,jet}}$ is associated with higher purities and higher efficiencies. After the respective correction and unfolding procedures, the momentum weight dimension is projected into the pair distributions of every $(R_{\textrm{L}}, p_{\mathrm{T,\,jet}})$ bin to obtain the EECs with their momentum weights unfolded. To corroborate the validity of the full correction procedure, a closure test is performed using simulation where statistical agreement within unity is obtained.

The systematic uncertainties are estimated by examining aspects such as the modeling of the detector response, the criteria for identifying charged hadrons reconstructed from unrelated hits, the unfolding procedure, the \Z-boson reconstruction efficiency, and the jet identification requirements. To obtain the total systematic uncertainty, the measurement is repeated by varying these aspects independently of each other; the deviations of such variations from the baseline values are added in quadrature. The total systematic uncertainty depends on the $p_{\mathrm{T,\,jet}}$ bin; it ranges between $6\%$ and $47\%$ for \mbox{$20<p_{\mathrm{T,\,jet}}<30\gev$}, $11\%$ and $27\%$ for \mbox{$30<p_{\mathrm{T,\,jet}}<50\gev$}, and $7\%$ and $21\%$ for \mbox{$50<p_{\mathrm{T,\,jet}}<100\gev$}. The main contributions are those related to the modeling of the detector response, the criteria to identify misreconstructed tracks, and the unfolding.

\begin{figure}[tb]
    \centering
    \includegraphics[width=\linewidth]{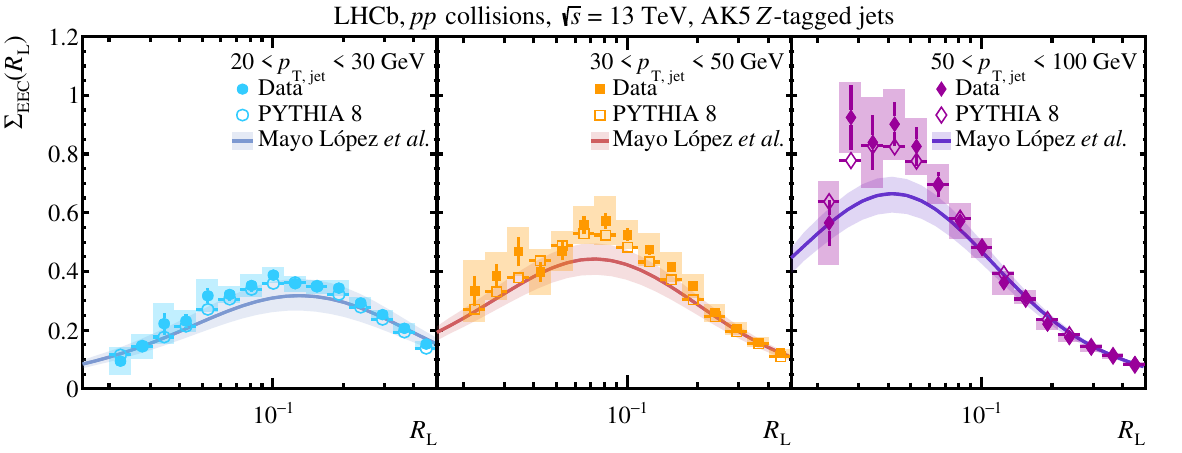}
    \caption{EECs as a function of $R_{\textrm{L}}$ for anti-$k_T$ jets with $R=0.5$ (AK5), in the indicated $p_{\mathrm{T,\,jet}}$ regions. For data, the vertical lines represent the statistical uncertainties and the shaded bands the total uncertainty. The theoretical predictions are based on Refs.~\cite{Barata:2024wsu,Barata:2026uby}.}
    \label{fig:eec}
\end{figure}

In Fig.~\ref{fig:eec}, the EECs as a function of the angular distance $R_{\textrm{L}}$ in bins of $p_{\mathrm{T,\,jet}}$ are shown. The transition region displays a shift to lower values of $R_{\textrm{L}}$ as $p_{\mathrm{T,\,jet}}$ increases, which is a behavior governed by $R_{\textrm{L}_{\textrm{transition}}} \sim \lqcd/ \langle p_{\mathrm{T,\,jet}}\rangle $~\cite{Komiske:2022enw}. The theory curves describe remarkably well the features of the data within uncertainties across the perturbative and nonperturbative regimes. An important aspect is the location of the peak, which in the theory is predominantly determined by a nonperturbative parameter extracted from \epem collisions; therefore, the proximity of the peak positions between data and theory supports universality of the transition region across the hadronization environments of \epem and \proton\proton collisions~\cite{Barata:2024wsu, Barata:2026uby}.

\begin{figure}[tb]
    \centering
    \includegraphics[width=0.7\linewidth]{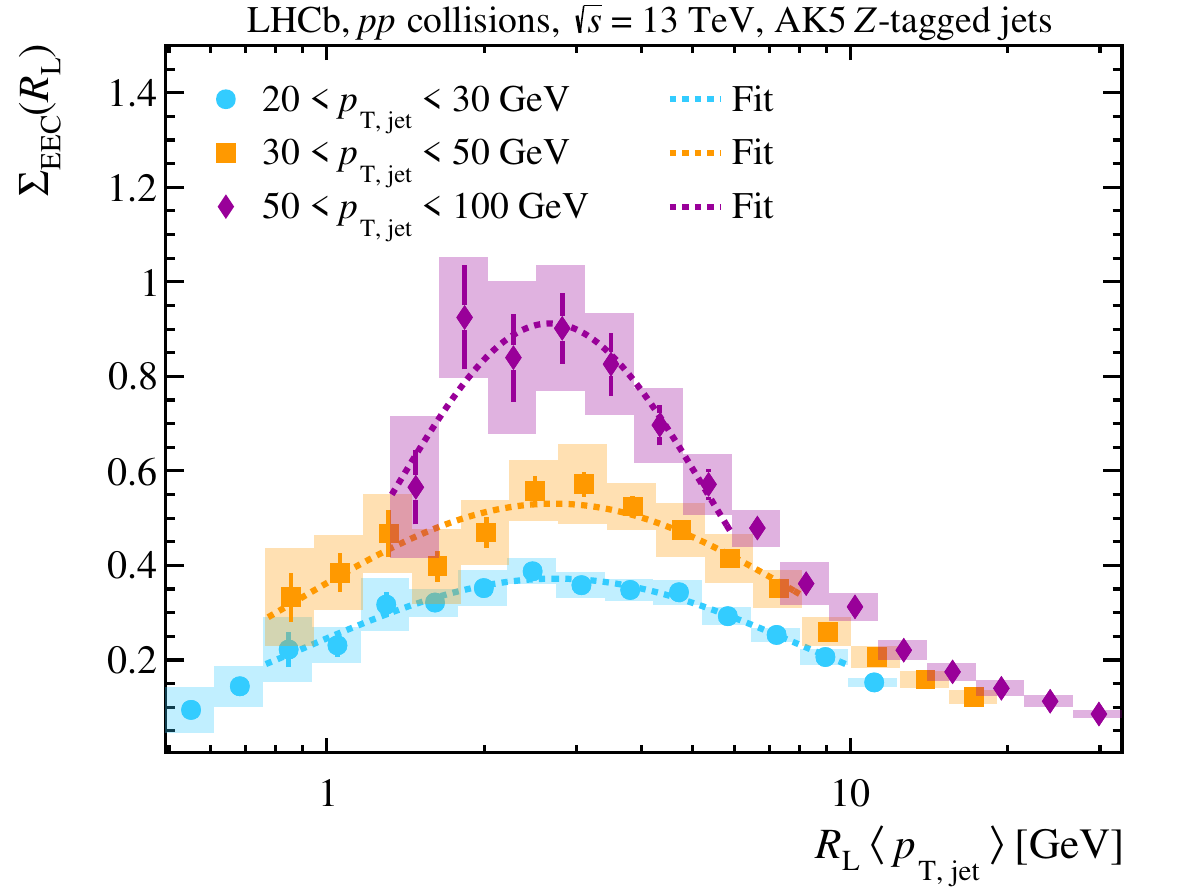}
    \caption{EECs as a function of $R_{\textrm{L}}\langle p_{\mathrm{T,\,jet}}\rangle$ for anti-$k_T$ jets with $R=0.5$ (AK5), in the indicated $p_{\mathrm{T,\,jet}}$ regions. The vertical lines represent the statistical uncertainties and the shaded bands the total uncertainty. The dashed lines show the fits performed to the distributions.}
    \label{fig:eec-rljetpt}
\end{figure}

In Fig.~\ref{fig:eec-rljetpt}, the EECs as a function of $R_{\textrm{L}}\langle p_{\mathrm{T,\,jet}}\rangle$ are shown for different ranges of $p_{\mathrm{T,\,jet}}$. The distributions highlight a universal scale governing the transition region for light-quark-initiated jets across $p_{\mathrm{T,\,jet}}$, as evidenced by the alignment of the transition regions in the data. The location of each transition region is estimated using fits performed over a symmetric region around the peak of each $p_{\mathrm{T,\,jet}}$ bin, and using a functional form inspired in previous studies~\cite{ALICE:2026zyx}. The extracted locations are $2.74\pm0.25\gev$ for \mbox{$20<p_{\mathrm{T,\,jet}}<30\gev$}, $2.71\pm0.50\gev$ for \mbox{$30<p_{\mathrm{T,\,jet}}<50\gev$}, and $2.67\pm0.22\gev$ for \mbox{$50<p_{\mathrm{T,\,jet}}<100\gev$}, where a systematic uncertainty is assigned by repeating the fits in different $R_{\textrm{L}}$ ranges. Therefore, the alignment is corroborated within uncertainties. 

\begin{figure}[bt]
    \centering
    \includegraphics[width=\linewidth]{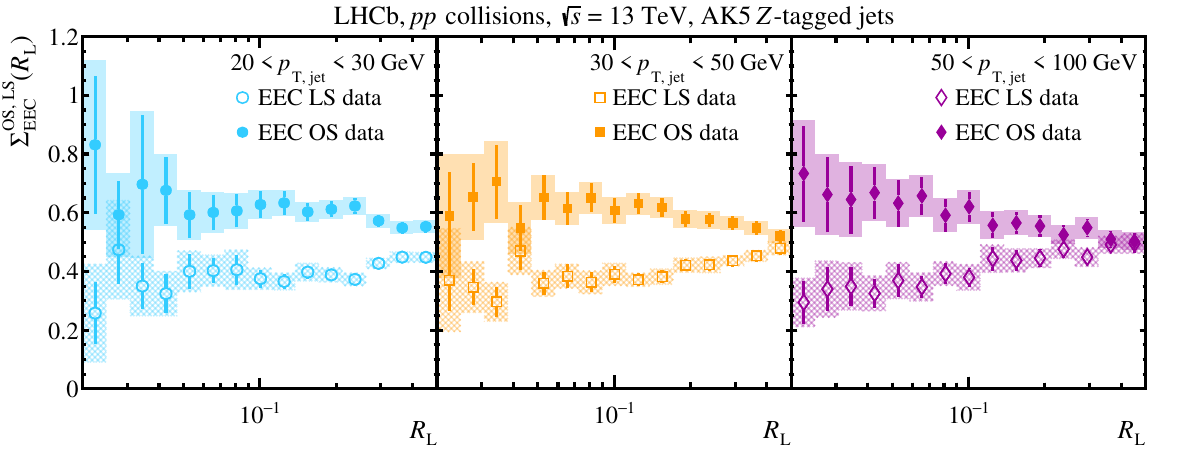}
    \caption{Like-sign and opposite-sign EECs as a function of $R_{\textrm{L}}$ for anti-$k_T$ jets with $R=0.5$ (AK5), in the indicated $p_{\mathrm{T,\,jet}}$ regions. The vertical lines represent the statistical uncertainties, and the shaded bands the total uncertainty.}
    \label{fig:eec-chargedep-sep}
\end{figure}

\begin{figure}[t]
    \centering
    \includegraphics[width=\linewidth]{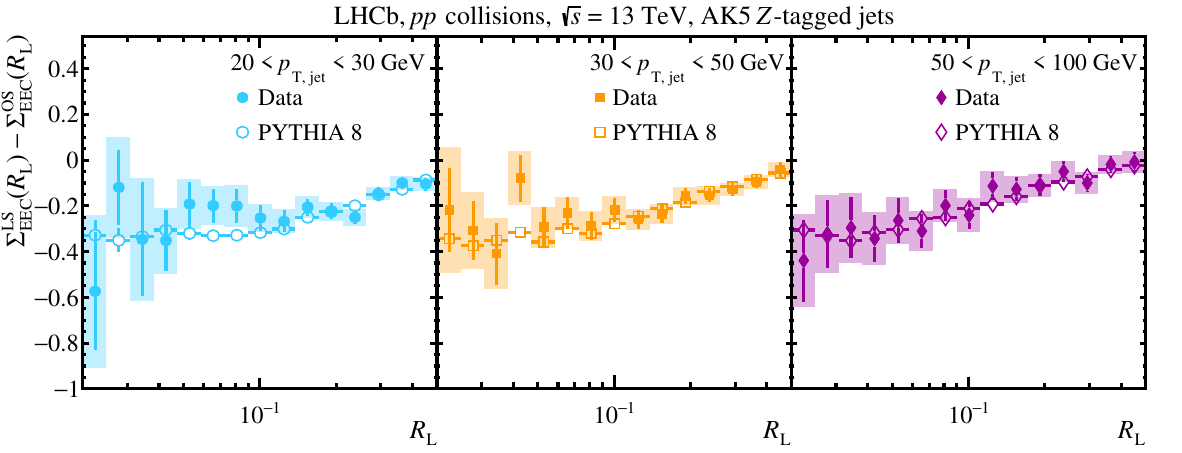}
    \caption{Charge-dependent EECs as a function of $R_{\textrm{L}}$ for anti-$k_T$ jets with $R=0.5$ (AK5), in the indicated $p_{\mathrm{T,\,jet}}$ regions. For data, the vertical lines represent the statistical uncertainties and the shaded bands the total uncertainty.}
    \label{fig:eec-chargedep}
\end{figure}

In Fig.~\ref{fig:eec-chargedep-sep}, the results for the like-sign and opposite-sign two-point energy correlators as a function of the angular distance $R_{\textrm{L}}$ are shown. The opposite-charge correlations dominate across most of the phase space, resulting in negative values of the charge-dependent correlator (see Fig.~\ref{fig:eec-chargedep}). It can be seen that the shape of these energy correlators is relatively constant in the free-hadron region for $p_{\mathrm{T,\,jet}}<50\gev$; in the case of $p_{\mathrm{T,\,jet}}>50\gev$, the free-hadron region is too small to make a conclusive statement. In the perturbative region, the correlations tend towards similar values as $R_{\textrm{L}}$ increases.

In Fig.~\ref{fig:eec-chargedep}, the differences for the charge-dependent two-point energy correlators as a function of the angular distance $R_{\textrm{L}}$ are shown. The negative values across the phase space highlight the dominance of the opposite-sign correlations. As $R_{\textrm{L}}$ decreases from $0.5$ to the transition-region values, the opposite-sign correlations are enhanced, a behavior that could be attributed to an emergent mechanism that is producing opposite-charged hadrons as angles get smaller. For instance, this could be in part due to a string-breaking-like mechanism, where opposite-charged pairs can be produced at small angles~\cite{Lee:2023npz}. The \pythia simulations describe the data well in this region, which could be taken as support for the previous supposition. In the free-hadron region, the distributions are consistent with an overall flat shape for the $p_{\mathrm{T,\,jet}}<50\gev$ range, in agreement with a recent measurement from the ALICE collaboration~\cite{ALICE:2026zyx}, but within uncertainties they are also consistent with the negative slope for $0.02 < R_{\textrm{L}} < 0.10$ observed by the STAR collaboration in their single reported $p_{\mathrm{T,\,jet}}$ bin with $20 < p_{\mathrm{T,\,jet}} < 30\gev$~\cite{STAR:2025jut}. 

In summary, this Letter reports the first measurements of inclusive, like-sign, opposite-sign, and charge-dependent EECs in \Z-tagged jets and at forward rapidities. The inclusive EECs display behavior consistent with that observed in previous measurements, such as the $p_{\mathrm{T,\,jet}}$-dependent shift of the transition region~\cite{CMS:2024mlf, ALICE:2024dfl, ALICE:2025igw, STAR:2025jut, CMS:2025ydi,ALICE:2026zyx}, and the alignment of the transition regions~\cite{ALICE:2024dfl, STAR:2025jut, ALICE:2026zyx}. The successful description of the data by phenomenological calculations supports the universality of the transition region across different collision systems, such as \epem and \proton\proton, in light-quark-initiated jets. Further higher-precision phenomenological studies will be useful to corroborate this observation. Moreover, the alignment of the transition region across different $p_{\mathrm{T,\,jet}}$ ranges supports a characteristic confinement scale for light-quark jets in \proton\proton collisions. The charge-dependent results display a behavior in the free-hadron region consistent with previous measurements with different flavor composition jets.  Further studies will be valuable to probe the charge-dependent dynamics in this region. Also, the enhancement of the opposite-sign correlation as $R_{\textrm{L}}$ decreases in the perturbative region indicates an emergent mechanism that produces opposite-charge pairs with small angular separation; alongside a good description of data by \pythia, this suggests a contribution from quark-antiquark production via a string-breaking mechanism.

The reported measurements show the robustness of energy correlators as a formalism to study jet-formation dynamics and probe the universality of nonperturbative effects in QCD. The light-quark-enriched jet sample used for this measurement provides a clean channel to study the behavior of hadronization in \proton\proton collisions; moreover, it establishes a baseline from which future studies can investigate the flavor dependence of EECs by comparing these results with other hadron-collider measurements where the fraction of light-quark jets is substantially smaller. These results open the door for future energy-correlator measurements at \lhcb, where the abundance of heavy-flavor jets offers a compelling opportunity to study how mass effects shape the perturbative and nonperturbative regimes of QCD.
\section*{Acknowledgements}
%
%
\noindent We thank Xoán Mayo López for providing theoretical predictions and João Barata, Zhong-Bo Kang, Jani Penttala, Ian Moult, Kyle Lee, Bianka Me\c{c}aj, and Evan Craft for helpful discussions. We express our gratitude to our colleagues in the CERN
accelerator departments for the excellent performance of the LHC. We
thank the technical and administrative staff at the LHCb
institutes.
We acknowledge support from CERN and from the national agencies:
ARC (Australia);
CAPES, CNPq, FAPERJ and FINEP (Brazil); 
MOST and NSFC (China); 
CNRS/IN2P3 and CEA (France);  
BMFTR, DFG and MPG (Germany);
NKFIH (Hungary);              
INFN (Italy); 
NWO (Netherlands); 
MNiSW and NCN (Poland); 
MEC/IFA (Romania); 
MICIU and AEI (Spain);
SNSF and SER (Switzerland); 
NASU (Ukraine); 
STFC (United Kingdom); 
DOE NP and NSF (USA).
We acknowledge the computing resources that are provided by ARDC (Australia), 
CBPF (Brazil),
CERN, 
IHEP and LZU (China),
IN2P3 (France), 
KIT and DESY (Germany), 
INFN (Italy), 
SURF (Netherlands),
Polish WLCG (Poland),
IFIN-HH (Romania), 
PIC (Spain), CSCS (Switzerland), 
GridPP (United Kingdom),
and NSF (USA).  
We are indebted to the communities behind the multiple open-source
software packages on which we depend.
Individual groups or members have received support from
RTP (Australia), 
FWO Odysseus grant G0ASD25N (Belgium), 
Key Research Program of Frontier Sciences of CAS, CAS PIFI, CAS CCEPP (China); 
Minciencias (Colombia);
EPLANET, Marie Sk\l{}odowska-Curie Actions, ERC and NextGenerationEU (European Union);
A*MIDEX, ANR, IPhU and Labex P2IO, and R\'{e}gion Auvergne-Rh\^{o}ne-Alpes (France);
Alexander-von-Humboldt Foundation (Germany);
ICSC (Italy); 
Severo Ochoa and Mar\'ia de Maeztu Units of Excellence, GVA, XuntaGal, GENCAT, InTalent-Inditex and Prog.~Atracci\'on Talento CM (Spain);
the Leverhulme Trust, the Royal Society and UKRI (United Kingdom).



\clearpage

\section*{End matter}
\label{sec:appendix}

\appendix

\section{Unfolding process}\label{appendix:unfolding}

\begin{figure}[h!]
    \centering
    \includegraphics[width=0.95\linewidth]{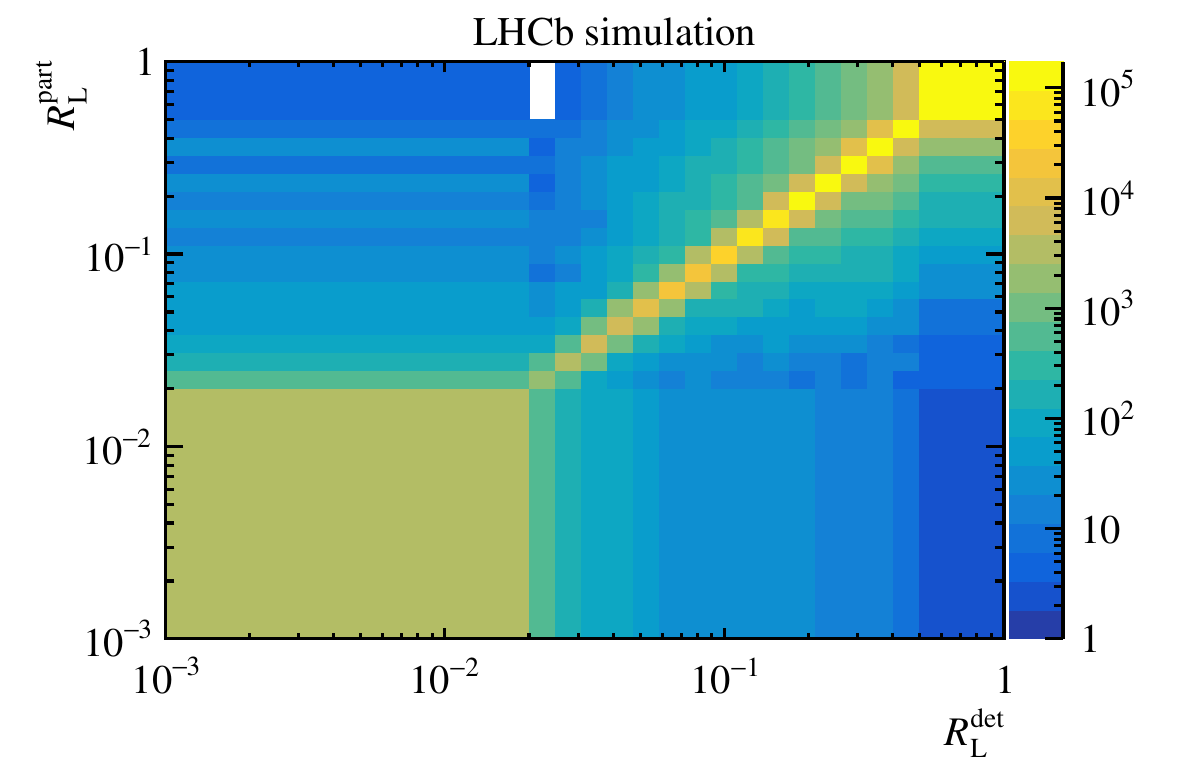}
    \caption{Response matrix for the angular distance. The horizontal axis displays the corresponding detector-level variable and the vertical axis the corresponding particle-level variable.}
    \label{fig:rm-rl}
\end{figure}

\begin{figure}[h!]
    \centering
    \includegraphics[width=0.95\linewidth]{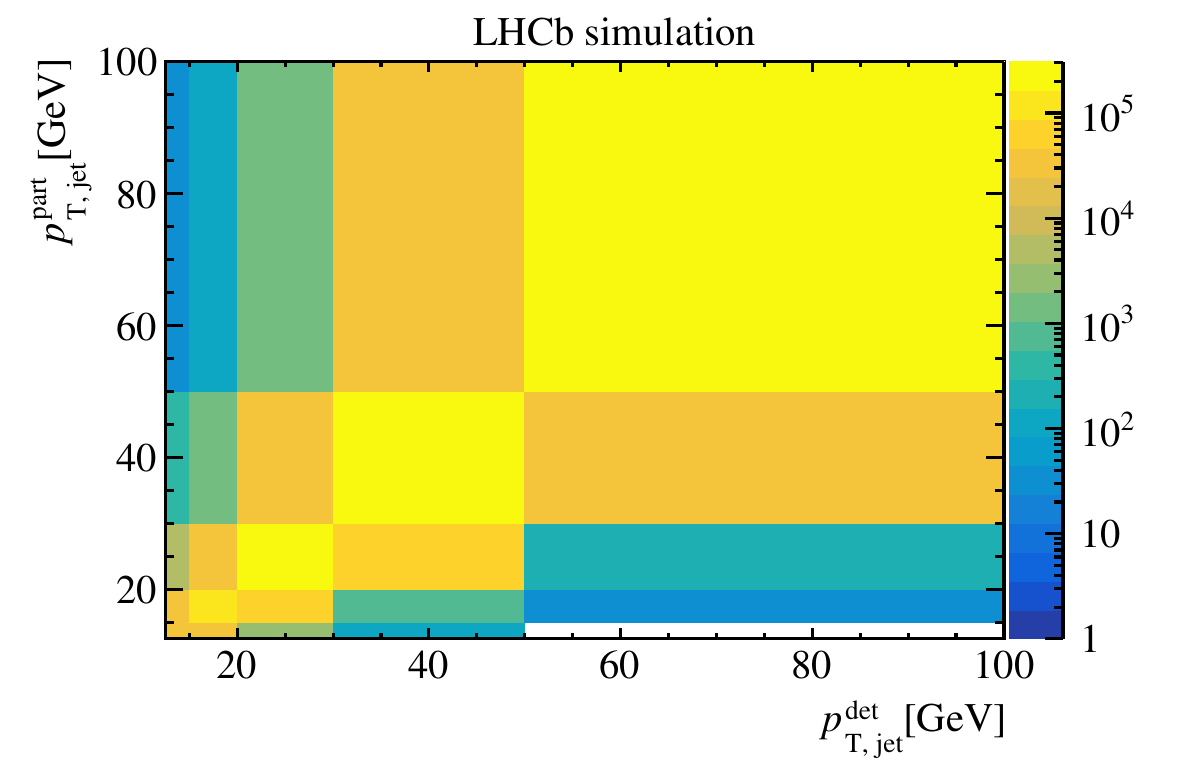}
    \caption{Response matrix for the transverse momentum of the jet. The horizontal axis displays the corresponding detector-level variable and the vertical axis the corresponding particle-level variable.}
    \label{fig:rm-jetpt}
\end{figure}

\begin{figure}[h!]
    \centering
    \includegraphics[width=0.95\linewidth]{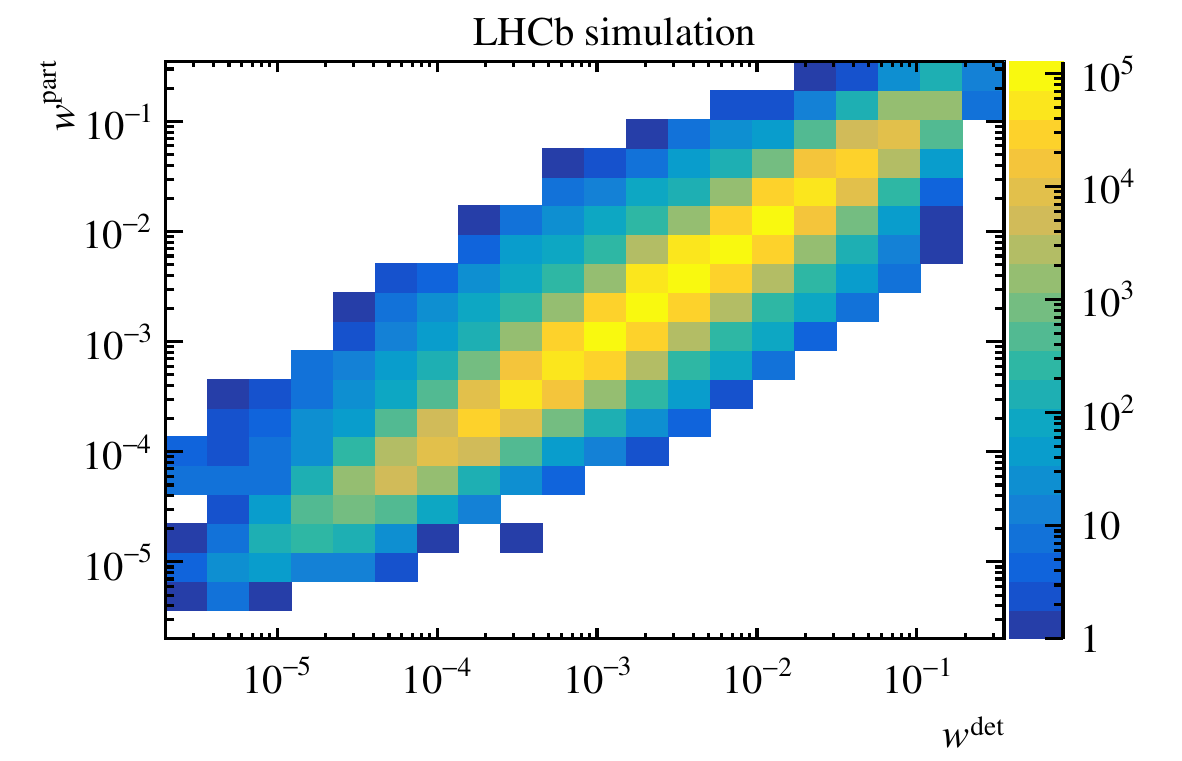}
    \caption{Response matrix for the momentum weights. The horizontal axis displays the corresponding detector-level variable and the vertical axis the corresponding particle-level variable.}
    \label{fig:rm-weight}
\end{figure}

In this Letter, a central part of the data treatment is the unfolding process. The unfolding is performed with the IBU algorithm~\cite{DAGOSTINI1995487} as implemented in the \textsc{RooUnfold} package~\cite{RooUnfold}, and simultaneously in three dimensions: $R_L$, $p_{\mathrm{T,\,jet}}$, and the momentum weights. Given the dimensionality of the unfolding, the response matrices are displayed in Figs.~\ref{fig:rm-rl},~\ref{fig:rm-jetpt}, and~\ref{fig:rm-weight}, for each unfolded variable on its own. The response matrices display a diagonal behavior. Although $R_L$ is smeared across the phase space, it also displays a diagonal behavior. The smearing observed in the response matrices points to the importance of the unfolding process in this measurement.
\clearpage


\addcontentsline{toc}{section}{References}
\ifx\mcitethebibliography\mciteundefinedmacro
\PackageError{LHCb.bst}{mciteplus.sty has not been loaded}
{This bibstyle requires the use of the mciteplus package.}\fi
\providecommand{\href}[2]{#2}

 
\newpage
\centerline
{\large\bf LHCb collaboration}
\begin
{flushleft}
\small
R.~Aaij$^{39}$\lhcborcid{0000-0003-0533-1952},
M.~Abdelfatah$^{71}$,
A.S.W.~Abdelmotteleb$^{59}$\lhcborcid{0000-0001-7905-0542},
C.~Abellan~Beteta$^{53}$\lhcborcid{0009-0009-0869-6798},
F.~Abudin\'en$^{61}$\lhcborcid{0000-0002-6737-3528},
T.~Ackernley$^{63}$\lhcborcid{0000-0002-5951-3498},
A.A.~Adefisoye$^{71}$\lhcborcid{0000-0003-2448-1550},
B.~Adeva$^{49}$\lhcborcid{0000-0001-9756-3712},
M.~Adinolfi$^{57}$\lhcborcid{0000-0002-1326-1264},
P.~Adlarson$^{87,44}$\lhcborcid{0000-0001-6280-3851}
% ...

{\footnotesize \it

% ...
$^{39}$Nikhef National Institute for Subatomic Physics, Amsterdam, Netherlands\\
% ...
$^{44}$National Center for Nuclear Research (NCBJ), Warsaw, Poland\\
% ...
$^{49}$Instituto Galego de F{\'\i}sica de Altas Enerx{\'\i}as (IGFAE), Universidade de Santiago de Compostela, Santiago de Compostela, Spain\\
% ...
$^{53}$Physik-Institut, Universit{\"a}t Z{\"u}rich, Z{\"u}rich, Switzerland\\
% ...
$^{57}$H.H. Wills Physics Laboratory, University of Bristol, Bristol, United Kingdom\\
% ...
$^{59}$Department of Physics, University of Warwick, Coventry, United Kingdom\\
% ...
$^{61}$School of Physics and Astronomy, University of Edinburgh, Edinburgh, United Kingdom\\
$^{62}$School of Physics and Astronomy, University of Glasgow, Glasgow, United Kingdom\\
$^{63}$Oliver Lodge Laboratory, University of Liverpool, Liverpool, United Kingdom\\
% ...
$^{71}$Syracuse University, Syracuse, NY, United States\\
% ...
$^{87}$Department of Physics and Astronomy, Uppsala University, Uppsala, Sweden, associated to $^{62}$\\
% ...
}
\end{flushleft}

\end{document}